# Single shot diagnosis of ion channel dysfunction from assimilation of cell membrane dynamics

Paul G. Morris,[1,2]† Joseph D. Taylor,[1]† Julian F.R. Paton,[3] Alain Nogaret[1]*

[1] Department of Physics, University of Bath, Claverton Down, Bath, United Kingdom.

[2] School of Physiology, Pharmacology and Neuroscience, University of Bristol, Bristol, United Kingdom.

[3] Manaaki Manawa – the Centre for Heart Research, Department of Physiology, Faculty of Medical and Health Sciences, University of Auckland, Grafton, Auckland, New Zealand.

†These authors contributed equally to this work

*Corresponding Author: A.R.Nogaret@bath.ac.uk

## Abstract

Many neurological diseases originate in the dysfunction of cellular ion channels. Their diagnosis presents a challenge especially when alterations in the complement of ion channels are a priori unknown. Current approaches based on voltage clamps lack the throughput necessary to identify the mutations causing changes in electrical activity. Here, we introduce a single-shot method for diagnosing changes in the complement of ion channels from changes in the electrical activity of a cell. We developed data assimilation (DA) to estimate the parameters of individual ion channels and from these parameters reconstruct the ionic currents of hippocampal CA1 neurons to within ±11% of their actual value. DA correctly predicts which ionic current is altered and by how much after we blocked the BK, SK, A and HCN channels with selective antagonists of known potency. We anticipate our assay will transform the treatment of neurological disease through comprehensive diagnosis and drug screening.

## Introduction

Channelopathies, in which certain ion channels are either absent or display abnormal conductances, occur congenitally or by autoimmune disease.[1–3] The complement of ion channels in a cell membrane underpins important aspects of neuronal function such as the shape of action potentials[4], adaptive versus non-adaptive firing response to an excitatory input[5], the integration of synaptic inputs[6,7] and some forms of short or long-term memory.[8] Therefore, the dysfunction of a single channel type can substantially alter cellular behaviour: for example, channelopathies are known to be the causative factor in many forms of epilepsy[9,10], pain disorders[1], cystic fibrosis[11], and cardiac arrhythmias[12], and form part of a more complex pathophysiology in Parkinson's[13,14] and Alzheimer's[15,16] diseases, Rett syndrome[17], and autism.[18] The need for efficient diagnosis of channelopathies has driven research on patch-sequencing that correlates gene expression with some morphological and electrical properties.[19] *In-vitro* electrophysiology remains the clinical benchmark, however this approach requires several pharmacological manipulations and is highly time consuming, particularly if the implicated channel is unknown. A single shot screening method is therefore needed that relates the changes in electrical properties of a cell to alterations in its ion channels.

In this work, we demonstrate a powerful diagnosis method based on statistical data assimilation (DA) that extracts information from current-clamp recordings to predict the alterations in ionic currents across the complement of ion channels of a neuron. The method synchronizes a Hodgkin-Huxley-like model[20] to oscillations of the membrane voltage to estimate ion channel parameters[21–24] such as maximal conductances, voltage thresholds, slopes of activation curves, and recovery times constants that constitute the fingerprints of individual ion channels.[25–27] The predictive power of our approach is based on the observation that ionic currents reconstructed from estimated parameters carry an uncertainty three times lower than the parameters themselves. We subsequently use the ionic charge transferred per action potential as a reliable metric to predict ion channel alterations induced by ion channel antagonists. We find that changes in predicted ionic charge match the selectivity and potency of well-characterized inhibitor compounds applied to block the BK, SK, A-type $K^+$, and HCN ion channels. This approach is to our knowledge unique in inferring changes across all ion channels from changes in dynamics of a cell membrane. Our method may be used to screen drugs targeting ion channel dysfunction and to diagnose channelopathies.

## Results

*Statistical Data Assimilation of pharmacologically altered neurons*

The statistical DA workflow is schematically depicted in **Figure 1**. We began by taking current-clamp measurements of hippocampal CA1 neurons, recording membrane voltage oscillations driven by a chaotic current waveform designed to elicit responses on multiple time scales, in both the hyperpolarized and the depolarized regimes (**Figure 1a**). Each neuron was recorded twice over a 2000ms long epoch: first in the natural state and a second time after applying an antagonist to block a specific ion channel (**Figure S1**). We then synchronized the neuron model to electrophysiological recordings using interior point optimization[28] within a constrained nonlinear optimization framework. The neuron model was a single compartment Hodgkin-Huxley-type system incorporating the 9 ion channels most prevalent in the CA1 soma (**Table S1**). Interior point optimization inferred the 67 parameters that best synchronize the model to electrophysiological data over a 800ms long assimilation window (**Table S2**). One set of 67 parameters was obtained from pre-drug data ($p_{pre}$) and another from post-drug data ($p_{post}$). Preliminary assimilations of model-generated data successfully recovered the 67 parameters of the original model to within 0.2% and with a 100% convergence rate.[21,30] Convergence was achieved irrespective of starting conditions and the positioning of the assimilation windows in the 2000ms long epoch. This indicates that the observability[31] and identifiability criteria[32] which are necessary to reconstruct the model's state variable from measurements, are fulfilled. In contrast, the problem of assimilating biological neuron data is complicated by our lack of knowledge of the exact model. Model error introduces correlations between some parameter estimates. As a result, parameter search tends to converge towards multiple solutions depending on the choice of starting conditions. In order to mitigate the uncertainty on parameters, we generated a statistical sample of parameters $p_{pre,1}$... $p_{pre,R}$ and $p_{post,1}$... $p_{post,R}$ (**Figure 1b**) by assimilating $R$ windows offset by 80ms from each other (Figures 1a). We then completed $2R$ conductance models by inserting the pre-drug and post-drug parameters in the model equations. The ionic current waveforms (**Figure 1c,d**) and membrane voltage oscillations (**Figure 1e**) were predicted by forward-integrating the stimulating protocol (Figure 1a) with pre-drug and post-drug completed models. We then numerically integrated the current waveform of each ion channel, to obtain the ionic charge transferred per action potential, pre-drug and post-drug. We repeated this process for the $R$ assimilation windows to generate a statistical sample of ionic charge transferred (Figure 1c). In the process we used one action potential common to all assimilation windows. The statistical distributions of ionic charges were plotted (Figure 1c) and analyzed (Mann-Whitney) to estimate the median and mean predicted inhibition for each channel. To ensure that our predictions are not affected by the firing frequency of neurons, which we found to be particularly sensitive to potassium channel inhibition, we calculated the charge transfer at the site of a single action potential instead of over the entire assimilation window. Ionic current waveforms were reconstructed and analyzed in parallel allowing all current alterations to be predicted in one shot (Figure 1d). Forward integration of the model also generated the predicted membrane voltage time series (Figure 1e). The good agreement between the experimentally observed and the predicted voltage provides an intermediate validation point of our method.

*Accuracy of current and parameter predictions*

The main challenge to inferring biologically relevant information from actual neurons as opposed to model data is to minimize the error introduced in the parameter field by model error and, to a lesser extent, measurement error.[31-33] In order to quantify the impact of model/data error, we calculated 100 sets of parameters by assimilating model data corrupted by 100 different realizations of white noise (**Figure 2**). The parameters that deviate significantly from their true values (**Figure 2a**) are few and mainly associated with gate recovery times $(t, \epsilon)$ (Table S2). In order to clarify the nature of parameter correlations, we calculated the 67×67 covariance matrix of this dataset (**Figure 2b**). We find that the covariance matrix exhibits a block structure whereby the correlations between parameters pertaining to the same ionic current are greater than those pertaining to different ionic currents. These findings suggest that the greater parameter correlations might compensate each other in the calculation of ionic currents. This underpins our key hypothesis that ionic currents might be estimated with a higher degree of confidence than their underlying parameters. A calculation of standard deviations of ionic currents and parameters over a range of noise levels (**Figure 2c**) validates this hypothesis by predicting a 3 times lower uncertainty on ionic currents. This finding allows us to focus on ionic current as a metric of ion channel alterations, and to validate the magnitude of alterations against the effect of antagonists of known selectivity and potency. **Figure 2d** plots the eigenvalues of the covariance matrix which measure the lengths of semi-axes of the data misfit ellipsoid. There are six outliers at the left which point to six principal directions along which parameters are very loosely constrained with $\frac{\Delta p}{p} \approx 100\%$. Along the 61 other principal directions $\frac{\Delta p}{p}$ varies between 7% and 0.001% confirming that most parameter estimates are well constrained as observed in Figure 2a. We now use these findings to predict the selectivity and potency of four ion channel antagonists applied to rodent hippocampal neurons.

*Predicting the alterations of ion channels induced by four antagonists in hippocampal neurons*

*BK channel blockade*

The analysis of neurons subjected to BK channel blocker iberiotoxin (IbTX; 100 nM; **Figure 3**; $R$=15 pre-drug and post-drug) predicted a *12.1% reduction* in median and 14.8% reduction in mean BK-mediated charge per action potential. This was the only statistical discovery across all channels in the drug-applied data (**Figure 3a**; U=25; q<0.001; median charge 29.4 nC.cm$^{-2}$ [pre-IbTX], 25.9 nC.cm$^{-2}$ [post-IbTX]). An increase in median 'leak' charge transfer of 63.9 % (mean 46.0 ± 25.8 %) was also observed, which did not meet the 1% false discovery rate threshold (q=0.016). There were no statistical discoveries for any other channels. This demonstrates that models constructed by DA correctly predict the selectivity of IbTX. **Figure 3b** predicts the reduction in charge transfer through the BK channel targeted by IbTX.

Identically driven action potentials measured pre-IbTX and post-IbTX (**Figure 3c**) are compared to the action potentials predicted from our pre-IbTX and post-IbTX models (**Figure 3d**). The model correctly predicts the reduction in afterhyperpolarization (fAHP) observed post-IbTX. BK current waveforms were also predicted by forward-integration of the pre-IbTX and post-IbTX conductance models (**Figure 3e**). The area under both waveforms yielded the drop in BK-mediated charge transfer plotted in Figure 3b.

*SK channel blockade*

Following application of the SK-specific channel blocker apamin (150 nM; **Figure 4**; $R$=18 pre-drug and post-drug), our model predicted a *100% reduction* in median (74.0% mean) SK-mediated charge transfer per action potential. This was the only statistical discovery across all channels in the spike-normalized data (**Figure 4a**; U=65; q<0.001; median charge 1.66 nC.cm$^{-2}$ [pre-apamin], 0.000 [post-apamin]). Our model thus correctly predicts that apamin is an antagonist of the SK channel. **Figure 4b** shows the predicted potency of apamin by plotting the reduction in SK-mediated charge transfer from the pre-apamin state to the post-apamin state. The identically driven action potentials measured pre-apamin and post-apamin (**Figure 4c**) are compared to the action potentials predicted by our conductance models (**Figure 4d**). The models correctly predict the reduction in medium afterhyperpolarization (mAHP) observed post-apamin in the tail end of the action potential. Forward-integration of the models also predicted the SK current waveforms at the site of an action potential pre- and post-apamin (Figure 4e). These waveforms were integrated in time to obtain the predicted amounts of SK-mediated charge transfer which were then plotted in Figure 4b.

*Kv channel blockade*

Following application of 4-Aminopyridine (4-AP) to block the voltage-gated potassium channels (300 μM; **Figure 5**; $R$=19 pre-drug; $R$=18 post-drug), our completed models predicted a *24.3% reduction* in median charge transfer (19.0% mean) mediated by A-type K$^+$ channels (**Figure 5a**; U=52; q<0.01; median charge 26.1 nC.cm$^{-2}$ [pre 4-AP], 19.7 nC.cm$^{-2}$ [post 4-AP]). In addition, the model predicts a 10.0% increase in median charge transfer (8.8% mean) through the BK-channel (U=73; q<0.01; median charge 41.2 nC.cm$^{-2}$ [pre 4-AP], 45.3 nC.cm$^{-2}$ [post 4-AP]); and a 4.4% reduction in median (3.0% mean) Ca$^{2+}$-mediated charge (U=79; q<0.01; median charge 9.65 nC.cm$^{-2}$ [pre 4-AP], 9.23 nC.cm$^{-2}$ [post 4-AP]).

**Figure 5b** predicts the reduction in charge transfer through the A-type K$^+$ channels targeted by 4-AP. Action potentials measured pre 4-AP and post 4-AP (**Figure 5c**) match the action potentials predicted by our pre 4-AP and post 4-AP models (**Figure 5d**). The model correctly predicts the widening of action potentials induced by 4-AP which follows from a slower AHP repolarization. **Figure 5e** plots the *predicted* A-type K$^+$ current waveforms elicited within the same action potential. The predicted current amplitude drops sharply in response to 4-AP. The K$^+$ charge amounts transferred per action potential are obtained by integrating the pre 4-AP and post 4-AP current waveforms and plotted in Figure 5b.

*HCN channel blockade*

We finally applied the ZD7288 antagonist to block the HCN channels (50 μM, **Figure 6**; *R*=19 pre-drug and post-drug). Our completed models predict a *100% reduction* in median HCN-mediated charge (85% mean) transferred across the full length of the assimilation window (**Figure 6a**; U=81; q<0.01; median charge 1.618 μC.cm$^{-2}$ [pre-ZD7288], 0.0 [post-ZD7288]) from the pre-drug state. In addition, our model predicts a 31.2% increase in median (25.1% mean) leak current (U=77; q<0.01; median charge 3.195 μC.cm$^{-2}$ [pre-ZD7288], 4.193 μC.cm$^{-2}$ [post-ZD7288]). These numbers represent the HCN charge amounts transferred across one 800ms long assimilation window rather than per action potential as above. This is because the HCN current contributes to subthreshold oscillations unlike the SK, BK and A-type currents which contributes to action potentials. **Figure 6b** predicts the total blockage of the HCN channel targeted by ZD7288. The membrane voltage response to a hyperpolarizing current step applied before and after ZD7288 (**Figure 6c**) is compared to the responses predicted by the pre-ZD7288 and post-ZD7288 models to the same current step (**Figure 6d**). The model correctly predicts the faster adaptation and the reduced amplitude of the membrane voltage change post-ZD7288.

In order to validate the diagnosis of data assimilation, we now compare the predicted changes in ionic charge transfer to the selectivity and potency of each ion channel antagonist determined by IC50 analysis. The results are summarized in **Table 1.** The predicted reductions in charge transfer are in good agreement with degree of inhibition expected in SK, BK, A-type and HCN. We further discuss below the inhibition of sub-types of the SK, BK, A, and HCN channels. Besides correctly identifying the selectivity of known antagonists, DA is sensitive enough to pick up correlations between ion channels driven by the modulation of reversal potentials[34] (Figure 6a) or compensation mechanisms[35] (Figure 5a). We also determined the degree of confidence in our predictions by computing the coefficient of variation (Table 1). The results consistently show a ±11% uncertainty on charge estimates.

**Discussion**

Our data have shown that DA predicts which ion channel is altered and by how much, demonstrating its potential as a single-shot diagnosis of channelopathies. We now discuss the two factors limiting its predictive accuracy. The first is model error and the impact of model overspecification in the presence of model error. The second is the presence of auxiliary sub-units functioning in parallel within each ion channel when DA predicts an aggregate contribution of all subunits.

Figures 3a-6a demonstrate the remarkable ability of DA to disentangle the contributions of 9 different ionic channels in the membrane voltage time series and to assign drug induced changes to the correct ion channel being blocked. The identification relies on the uniqueness of the mathematical equations of each ionic current (Table S1). It is remarkable that the correct ion channel is still identified even when we know, from the existence of parameter correlations, that model equations are only an approximation of biological reality. There are, however, reasons to believe that any model error is residual because the completed models make excellent predictions of the membrane voltage pre-drug and post-drug (Figures 1e, 3d-6d). In order to evaluate the effect of model error on current estimates, we have deliberately introduced an erroneous gate exponent in the sodium current (NaT), changing the gate exponent $m^3h$ to $m^2h$. We find that the current waveforms estimated with the wrong model deviate only by a few percent from their true shape. We also find any drop in sodium current (NaT) induced by model error is compensated by a drop in potassium (A) current. This shows that a slightly wrong model still retains its ability to discriminate ion channels as observed in Figures 3a-6a. We have also added a supernumerary ion channel (muscarinic current) to the 9-channel erroneous model to investigate whether DA might assign a finite conductance to this channel to compensate for the erroneous $m^2h$ exponent of the sodium current. We find that no such compensation occurs. Instead, DA correctly filters out the supernumerary current assigning a vanishingly small conductance value to it, whether the NaT exponent is $m^3h$ or $m^2h$. This implies that larger models incorporating more ion channels than necessary are still able to disentangle the contributions of different ionic currents. This is important as larger models have the advantage of not requiring any prior assumptions as to which ion channel must be included in and would apply to a wide range of biological neurons. As the model size grows, simulations on model data ought to verify the stimulation protocol still satisfies identifiability criteria[31,32] as a prerequisite. Multicompartment models may also be used[22] however single compartment models have by and large been sufficiently detailed to accurately predict voltages and currents.[22-27] Assuming electrophysiological data included dendritic transmission line delays, these delays would be the same pre-drug and post-drug and therefore would not affect the predicted charge alterations reported in Figures 3a-6a.

We now discuss the predicted channel block in relation to the different contributions of auxiliary sub-units functioning in parallel within each ion channel.

*Prediction of BK channel alterations*

Our inference method correctly identifies a reduction in the BK-mediated current in response to the application of IbTX. The 12.1% median reduction in the BK current was the only statistical discovery validated by the Mann-Whitney criterion (q<1%) among all 7 ion channels analyzed (Figures 3a,3b). DA thus recognizes the high selectivity of IbTX which is a highly selective inhibitor of BK channels in hippocampal CA1 neurons.[36,37] BK channels have a very high unitary conductance.[38,39] They contribute to both the repolarization of the action potential, and to the fast period of afterhyperpolarization, (fAHP) as seen in Figure 3c following the application of IbTX. Our model correctly predicts the contribution of the BK channel to the repolarization and fAHP phases in addition to the changes induced by IbTX (Figure 3d). This is further validation that DA transfers biological relevant information to the complete model.

The response of the BK channel to IbTX is heavily modulated by the presence of the pore forming α subunit and auxiliary subunits (β1-β4)[37,40,41]. Our prediction of a 12.1% median reduction in the BK current (**Table 1**) is likely to aggregate the contributions of up to four sub-units. Generally, β1 and β3 do not appear to be expressed in the brain. β2 is highly expressed in astrocytes, and β4 is expressed in neurons. It has been suggested that a complement of the four β subunits (1:1 stoichiometry) may be required to confer full IbTX resistance[42]: channels with less than four β subunits would exhibit toxin sensitivity similar to channels totally lacking β4 subunits[42]. As stoichiometry is unknown in these neurons, a mix of configurations would result in the partial inhibition of BK-mediated currents by IbTX. The potency of IbTX has previously been evaluated for several configurations of β subunits[43] and is listed in **Table 1.** A close match to our 12.1% predicted degree of block is 12% expected block of the [α+β$_1$] combination.

*Prediction of SK channel alterations*

SK channels have a major role in the generation of AHP. Our predictions from recordings made using the highly specific SK blocker, apamin,[44] showed a 100% reduction in SK-mediated current (Figures 4a,4b). DA did not predict any notable attenuation in charge transfer across any other type of ion channel. This validated the predictive power of our inference method against the specificity of inhibition by apamin.

Apamin is a highly selective inhibitor of SK2 and SK3 channels, which mediate medium AHP currents (mI$_{AHP}$) with a relatively fast inactivation and decay[44]. Whilst apamin-sensitive I$_{AHP}$ currents have been shown to be present in the CA1 soma, their blockade is often masked by the activity of other voltage-gated potassium channels[45,46]. The majority of SK channel subunits in CA1 neurons are SK2[47,48], with SK3 showing relatively low expression, and SK1 being expressed in moderate levels[49] and being apamin insensitive.[50,51] At 150 nM, apamin is expected to completely block SK2/3-mediated mAHP[49]. Our prediction of a 100% block (Figure 4b) is therefore in excellent agreement with 100% block expected with apamin.

*Prediction of A-type and K channel alterations*

4-aminopyridine was applied to inhibit the voltage-dependent $K^+$ channels. These are accounted for by the A-type and K-delayed rectifier channels in our model. The A-type channels are known to be present in CA1 neurons[6] where they give fast activating and fast inactivating $K^+$ currents whose role is to suppress excitatory postsynaptic potentials and delay action potentials[52]. A-type $K^+$ currents flow through either the Kv1.4 or Kv4.2 channels[53], with Kv4.2 being more abundant[54,55]. Our prediction of a 24.3% median (19.0% mean) current reduction through the A-type $K^+$ channels (Figures 5a,5b) closely matches the 13-25% inhibition of A-currents expected from 4-AP, at 300µM concentration (Table 1)[56,57].

The K-delayed-rectifier channels include the Kv1-3,5 and 6 subfamilies which are also inhibited by 4-AP, with an IC50 of 200-1500 µM[58,59]. We expected to see a reduction in predicted activity for the K-channel; however, whilst a reduction was clearly visible in 4-AP (Figure 5a) this did not generate a discovery in the Mann-Whitney test. This is because the amount of charge transferred in the natural state is very low compared to other channels (NaT, A and BK). To statistically confirm the K-channel block, a larger sample of parameter estimates ($R \gg 19$) would be needed to reduce the variance on the predicted charge transfer.

*Prediction of HCN channel alterations*

Hyperpolarization-activated and cyclic nucleotide-gated (HCN) channels belong to the superfamily of voltage-gated pore loop channels. They are unique in possessing a reverse voltage-dependence that leads to activation upon hyperpolarization.[60] The HCN1 and HCN2 subunits are the most abundant in CA1 neurons,[61–63] and both subunits are represented in our model of the HCN channel. Under ZD7288, the predicted HCN current was reduced by 100% median (85.5% mean) (Figures 6a,6b). This result is a reasonably good match to the degree of block expected from previous work on CA1 neurons indicating a 70-85% mean reduction in HCN-mediated current.[64,65] Similar values were obtained from specific studies of the HCN1[66] and HCN2[67] subunits (Table 1).

*Prediction of secondary effects of pharmacological inhibition in other channels*

DA also predicted alterations in ion channels not specifically targeted by the 4-AP and ZD7288 antagonists (Figures 5a, 6a). The observation of collateral alterations is consistent with the modification of the electrochemical driving force by the antagonist which alters current flow through other ion channels. This mechanism is particularly effective at times when the blocked channel would otherwise have been activated. For example, a reduction in $K^+$ permeability during the afterhyperpolarization phase will change the electrochemical driving force of other ions during that period. The driving force of $Cl^-$ into the cell will increase whereas the $Na^+$ driving force will be reduced. In addition, potassium current through the BK channel can compensate for the blocked Kv channels and vice-versa.[68].

The co-lateral effect of IbTX is to increase the leak current as predicted in Figure 3a. This effect, which falls below the 1 % false discovery rate threshold, is likely to be caused by the increase in $K^+$ permeability when the large conductance of the BK channel is blocked. It is also notable that within the 4-AP dataset (Figure 5a), the BK current increases when the A-type channel is blocked. This is a well characterized effect of 4-AP which causes a persistent $K^+$ current[68] and increases the spike width[69,70] (Figure 5c). Our model correctly predicts this spike broadening (Figure 5d). DA also has sufficient sensitivity to pick up the second order increase in BK current when the A channel is blocked (Figure 5a). A small (4.4 %) reduction in median $Ca^{2+}$ channel current was also predicted in Figure 5. This is likely to result from a reduction in the electrochemical driving force on $Ca^{2+}$ caused by the decreased potassium permeability following each action potential. Lastly, when inhibiting HCN channels using ZD7288, DA predicted an increase in leak current (Figure 6a). Because ZD7288 also strongly inhibits sodium channels[71], the reduction in $Na^+$ permeability would increase $Cl^-$ efflux as observed in our results.

DA inference is expected to transform the diagnosis of neurological disease by providing an unbiassed quantitative assessment of ion channel dysfunction across the complement of ion channels. Its ability to predict the selectivity and potency of antagonists, together with second order correlations between ion channels, show that predictions are accurate and reliable. Our proof-of-concept will easily be extended to probing the ion channel mutations in neurological disease. Our approach removes the need to make educated guesses about which gene is responsible for any given disease and to design genetically modified animal models that could fail to translate to the human form of the disease. A case in kind are genetically modified mice models of Alzheimer's disease which have proven to be ill-suited as a model of the human form of the disease.[15] With our approach, functionally relevant genetic mutations can now be inferred directly from human cells without the need to develop an animal model which may potentially be flawed. Our method also fills a blind spot in transcriptomics and proteomics sequencing. These bottom-up sequencing methods do not discriminate between alterations which are relevant to electrical function from others which are not. The merit of DA is that it is a top-down method starting from the electrical output of a cell to infer only those ion channel alterations which are expressed and functionally relevant. This constitutes a breakthrough for screening new drugs and understanding the effect of drug binding to ion channels. Combined with its high throughput, our approach is attractive as an assay for drug modulation of ion channel activity.

In summary, the present study has reliably reconstructed the current waveforms of all ionic channels of a neuron by assimilating the information contained in the oscillations of its membrane voltage. We have predicted the alterations in ionic currents in response to four antagonists and found these predictions to be sufficiently accurate to match the selectivity and nominal potency of each antagonist, in addition to predicting second order correlations between ionic currents. Our simulations have further determined how prediction accuracy and reliability was achieved in spite of the approximate nature of the Hodgkin-Huxley systems and how the approach will be extended to other types of neurons. Our data assimilation method goes beyond state-of-the-art by providing rapid, unbiassed and comprehensive diagnosis of ion channel alterations from simple current-clamp measurements. The method has potential both as a diagnosis tool for channelopathies in disease, or as a screening device for drugs modulating ion channel function, thereby allowing a rapid identification of targets for clinical study.

## Materials and Methods

*Current clamp electrophysiology*

We recorded CA1 hippocampal neurons with a current clamp amplifier (Molecular devices, MultiClamp 700B). The amplifier input was driven by a LabView controller (National Instruments) interfaced with a National Instruments USB-6363 DAQ card which delivered the protocol signal to the amplifier and recorded the membrane voltage returned by the neuron. Prior to each series of experiments, the gain of the amplifier was adjusted to elicit a maximum number of action potentials per measurement epoch without causing depolarization block with excessive current amplitudes. The calibration protocol is described in **Figure S2**. The injected current protocols were designed to fulfil the identifiability criterion of the inverse problem, that is to excite the full dynamic range of the neuron. It comprised a mixture of hyperpolarizing and depolarizing current steps of different amplitudes and durations, and chaotic oscillations generated by the Lorenz96 system. Both the current stimulus and the membrane voltage were sampled at a rate of 100kHz. This time resolution gave 20 datapoints per action potential which is sufficient for interpolating the finer features of the neuron response.

Whole-cell current-clamp recordings were performed in acute brain slices from male Han Wistar rats at P15–17. Following decapitation, the brain was removed and placed into an ice-cold slicing solution composed of (mM): NaCl 52.5; sucrose 100; glucose 25; $NaHCO_3$ 25; KCl 2.5; $CaCl_2$ 1; $MgSO_4$ 5; $NaH_2PO_4$ 1.25; kynurenic acid 0.1, and carbogenated using 95% $O_2$/5% $CO_2$. A Campden 7000 smz tissue slicer (Campden Instruments UK) was used to prepare transverse hippocampal slices at 350 μm, which were then transferred to a submersion chamber containing carbogenated artificial cerebrospinal fluid (aCSF) composed of (mM): NaCl 124; glucose 30; $NaHCO_3$ 25; KCl 3; $CaCl_2$ 2; $MgSO_4$ 1; $NaH_2PO_4$ 0.4 and incubated at 30 °C for 1–5 h prior to use. Synaptic transmission was inhibited pharmacologically in order to remove random synaptic inputs from the surrounding network. To this end all experiments were performed in the presence of (μM) kynurenate 3, picrotoxin 0.05, and strychnine 0.01, to inhibit ionotropic glutamatergic, γ-aminobutyric acid (GABA)-ergic, and glycinergic neurotransmission respectively

For patching, slices were transferred to the stage of an Axioskop 2 upright microscope (Carl Zeiss) and pyramidal CA1 neurons were identified morphologically and by location using differential interference contrast optics. The chamber was perfused with carbogenated aCSF (composition as above) at 2 ml $min^{-1}$ at 30 ± 1 °C. Patch pipettes were pulled from standard walled borosilicate glass (GC150F, Warner Instruments) to a resistance of 2.5–4 MΩ, and filled with an intracellular solution composed of (mM): potassium gluconate 130; sodium gluconate 5, HEPES 10; $CaCl_2$ 1.5; sodium phosphocreatine 4; Mg-ATP 4; Na-GTP 0.3; pH 7.3; filtered at 0.2 µm.

Inhibitor compounds were selected for the predictability of their effects on ion channel types known to be present in hippocampal pyramidal neurons:

- SK channels were inhibited using apamin (150 nM);
- BK channels were inhibited with iberiotoxin (100 nM);
- HCN channels were inhibited with ZD7288 (50 µM);
- A and K channels were inhibited using 4-AP (300 µM).

The potency of each drug was obtained from IC50 values tabulated in the literature (**Table 1**), which we compared to the reduction in ionic charge transfer predicted by our DA method.

*Model description*

A single-compartment model of the CA1 pyramidal neurons was built using a conductance-based framework incorporating seven active ionic currents identified in the physiological literature as being prevalent in the soma of CA1 neurons[61,65,72], in addition to a voltage-independent leak current[73,74]. The complement of ionic channels includes transient sodium (NaT), persistent sodium (NaP), delayed-rectifier potassium (K), A-type potassium (A), low threshold calcium (Ca), large- and small-conductance $Ca^{2+}$-activated potassium (BK and SK respectively), and the hyperpolarization-activated cation channel (HCN). The density of calcium channels in the soma of CA1 neurons is much lower than in distal dendrites[75], however the internal $Ca^{2+}$ concentration activates the transfer of $K^+$ ions through the Ca-dependent BK and SK channels. Therefore our model equations need to include the calcium current. The equation of motion for the membrane voltage is:

$$C\frac{dV(t)}{dt} = -J_{NaT} - J_{NaP} - J_K - J_A - J_{Ca} - J_{BK} - J_{SK} - J_{HCN} - J_{Leak} + I_{inj}(t)/A \,, \quad (1)$$

where $C$ is the membrane capacitance, $V$ is the membrane potential, $I_{inj}(t)$ is the injected current protocol (Figure 1a), $A$ is the surface area of the soma, and $J_{NaT} \ldots J_{Leak}$ are the ionic current densities across the cell membrane. The equations describing individual ionic currents are given in **Table S1**. These currents depend on maximum ionic conductances ($g_{NaT}$, $g_K$, $g_{HCN}$...), reversal potentials ($E_{Na}$, $E_K$, $E_{HCN}$...), and gating variables ($m$, $h$, $n$, $p$, ...). The kinetics of each ionic gate is described by a first order equation and each gate activates (inactivates) according to a sigmoidal function of the membrane voltage. The equations for each ion channel are as follows:

*Sodium Channels* – The *activation* gate variables of the NaT and NaP channels were respectively:

$$m_\infty(V) = 0.5\left[1 + \tanh\left(\frac{V-V_m}{\delta V_m}\right)\right], \quad (2)$$

$$p_\infty(V) = 0.5\left[1 + \tanh\left(\frac{V-V_p}{\delta V_p}\right)\right], \quad (3)$$

where $V_m$, $V_p$ are the activation thresholds and $\delta V_m$, $\delta V_p$ are the widths of the gate transition from the open to the closed state. The activation time of NaT and NaP being very rapid (~0.1ms)[73] compared to other channels we have assumed it to be instantaneous. This simplification reduces model complexity and improves parameter identifiability in DA.

The kinetics of the NaT *inactivation* gate is given by :

$$\frac{dh(V,t)}{dt} = \frac{h_\infty(V) - h(V,t)}{\tau_h(V)}, \tag{4}$$

where the steady-state inactivation curve is:

$$h_\infty(V) = 0.5\left[1 + \tanh\left(\frac{V-V_h}{\delta V_h}\right)\right], \tag{5}$$

and the recovery time depends on the membrane voltage as:

$$\tau_h(V) = t_h + \epsilon_h\left[1 - \tanh^2\left(\frac{V-V_h}{\delta V_{\tau h}}\right)\right]. \tag{6}$$

$V_h$ is the inactivation threshold, $\delta V_h$ the width of the open/close transition of the inactivation gate. $t_h$ is recovery time away from the depolarization threshold and $t_h + \epsilon_h$ the recovery time at the depolarization threshold. $\delta V_{\tau h}$ is the width of the peak at half maximum.

*Potassium channels* - the non-inactivating delayed-rectifier current (K) and the rapidly inactivating A-type potassium current (A) have the form[76] given in **Table S1**. The kinetics of the A-type activation gate is:

$$\frac{da(V,t)}{dt} = \frac{a_\infty(V) - a(V,t)}{\tau_a(V)}, \tag{7}$$

where $a_\infty(V)$ and $\tau_a(V)$ are given by Eqs 5, 6 where the subscript $h$ is replaced with $a$ (Table S2).

The inactivation kinetics of the K and A-type channels are respectively given by:

$$\frac{dn(V,t)}{dt} = \frac{n_\infty(V) - n(V,t)}{\tau_n(V)}, \tag{8}$$

$$\frac{db(V,t)}{dt} = \frac{b_\infty(V) - b(V,t)}{\tau_b(V)}, \tag{9}$$

where $n_\infty(V)$, $\tau_n(V)$ and $b_\infty(V)$ and $\tau_b(V)$ are given by Eqs 5, 6 with the appropriate substitution of indices (Table S2).

*Calcium activated potassium channels* – The BK and SK current are Ca²⁺ activated potassium currents found in the soma of hippocampal pyramidal cells[75]. The BK current is sensitive to both membrane voltage and internal Ca²⁺ concentration whereas the SK current only depends on the Ca²⁺ concentration (Table S1). Both currents are dependent of the internal calcium concentration given by[76]:

$$\frac{d[Ca]_{in}}{dt} = \frac{[Ca]_\infty - [Ca]_{in}}{\tau_{ca}} - \frac{J_{Ca}}{2z}, \qquad (10)$$

where $[Ca]_\infty$ is the equilibrium concentration, $\tau_{ca}$ is the recovery time, $z$ is Faraday's constant and $J_{ca}$ is the calcium current whose expression is given in Table S1. The calcium current had voltage-dependent activation and inactivation gates, *s* and *r*, respectively.[76] The kinetics and activation curves of *s* and *r* are given by Eqs.4-6 where subscript *h* is replaced with the *s* and *r* subscripts of the Ca parameters (Table S2).

The BK current has two gate variables *c*, *d* while the SK channel has one *w*. The form of the ultrafast SK activation gate, *w*, is given by Warman et al.[77] as:

$$w \equiv w_\infty([Ca]_{in}) = 0.5\left[1 + tanh\left\{\left\{V - V_w + 130\left\{1 + \tanh\left(\frac{[Ca]_{in}}{0.2}\right)\right\} - 250\right\}/\delta V_w\right\}\right] \qquad (11)$$

The slower activation gate of the BK channel, c, follows a first order rate equation:

$$\frac{d\,c}{dt} = \frac{c_\infty(V,[Ca]_{in}) - c}{\tau_c} \qquad (12)$$

with a steady-state activation curve given by:

$$c_\infty(V,[Ca]_{in}) = 0.5\left[1 + tanh\left\{\left\{V - V_c + 130\left\{1 + \tanh\left(\frac{[Ca]_{in}}{0.2}\right)\right\} - 250\right\}/\delta V_c\right\}\right] \qquad (13)$$

The inactivation gate of the BK channel, d, similarly follows a first order rate equation:

$$\frac{d\,d}{dt} = \frac{d_\infty(V,[Ca]_{in}) - d}{\tau_d(V)} \qquad (14)$$

with

$$d_\infty(V, [Ca]_{in}) = 0.5 \left[1 + tanh\left\{\left\{V - V_d + 130\left\{1 + tanh\left(\frac{[Ca]_{in}}{0.2}\right)\right\} - 250\right\}/\delta V_d\right\}\right] \quad (15)$$

$$\tau_d(V) = t_d + \epsilon_d \left[1 - tanh^2\left(\frac{V-V_d}{\delta V_{\tau d}}\right)\right]. \quad (16)$$

The existence of the SK and BK ionic currents was validated by much improved fits of the height and shape of action potentials, and their AHP region (Figure 1e). Without the SK and BK currents, the model clips action potentials at 80% of their maximum height. In total, our conductance model had the 67 adjustment parameters listed in Table S2.

*Parameter estimation and current prediction*

Our interior point method optimizes the parameter vector $\boldsymbol{p}^*$ and the initial state vector $\mathbf{x}^*(t=0)$ by minimizing the misfit between the experimental membrane voltage, $V_{data}$, and the membrane voltage variable, $V$, at each time point $t_i$ ($i = 0, \ldots, N$) of the assimilation window (Figure 1a). This misfit is evaluated by the least-square cost function:

$$C(p, x(0)) = \frac{1}{2} \sum_{i=0}^{N} \left\{[V_{data}(t_i) - V(t_i, p, x(0))]^2 + u(t_i)^2\right\}. \quad (17)$$

Minimization is done under both equality and inequality constraints. The equality constraints are the model equations (Eqs.1-16) linearized at each time point $t_i$. The inequality constraints are given by the lower and upper boundaries of the parameter search range, LB and UB, in Table S2. These are set by the user. The 67 parameter components of the parameter vector $\boldsymbol{p}^*$ are listed in Table S2. The state vector has 14 variables, $x(t) \equiv \{V(t), m(t), h(t), p(t), n(t), a(t), b(t), s(t), r(t), c(t), d(t), w(t), z(t), [Ca]_{in}\}$ that hold membrane voltage, gate variables, and internal calcium concentration. We used the interior-point-optimization algorithm developed by Wächter and Biegler[28] to minimize the cost function. Our assimilation windows were all 800ms long and were meshed by $N=40,000$ intervals of equal duration. We have reported the procedure for linearizing constraints in earlier publications[21,30] which we refer the reader to.

In order to stabilize the convergence of the parameter search, a control term $u(t_i)[V_{data}(t_i)-V(t_i)]$ was added to the right-hand side of Eq. 1[21,78] and as $u^2(t_i)$ in Eq.17. In well-posed assimilation problems, the $u(t_0)$ … $u(t_N)$ uniformly tends to zero as $p$ converges to the solution $p^*$. The model error and experimental error encountered with biological neurons makes the problem ill-posed. Model error introduces correlations between some parameters which take multi-valued solutions when the initial guesses on state variables, parameters or data intervals vary. In this case the $u(t_i)$ also converge to zero except at times that coincide with action potentials. Models configured with optimal parameters that include a small subset of correlated parameters reliably predict membrane voltage oscillations and ionic current waveforms for a wide range of current injection protocols (**Figure 1, S1**). When the $u(t_i)$ failed to converge, the estimated parameters were discarded from the statistical analysis of the ion channels. Models configured with such parameters were unable to predict any sensible quantity. Prior to the analysis of biological recordings, we verified that our current protocol and DA procedure fulfilled the conditions of observability and identifiability on model data. These preliminary studies showed that DA recovered *all* 67 parameters to within 0.1% of their original value in the model used to produce the assimilated data. We verified the uniqueness and accuracy of solutions using the $R=19$ assimilation windows offset by 80ms (Figure 1a) and varying the starting values of $p^*$ and $x^*$.

The predicted ionic currents and membrane voltages were generated by forward integration of each completed model over the 2000ms long epoch, both pre- and post-drug. Current waveforms were integrated to obtain the total charge transferred through each channel in that epoch (Figures 3e, 4e, 5e). In order to eliminate the dependence on the neuron firing frequency, we divided the total charge transferred across the epoch by the number of action potentials to obtain the net charge transferred per spike, per ion channel (Figures 3a, 4a, 5a). To verify the predicted changes are not affected, we also plotted the changes in *total* charge transferred over the epoch both pre- and post-drug (**Figure S3**). We verify that both methods gave similar results with small differences arising from sub-threshold current flow between action potentials.

The model equations were differentiated symbolically using our custom-built Python library pyDSI to generate the C++ code of the optimization problem.[22–24] This code was then inserted in the open-source IPOPT software [www.coin-or-org/ipopt] implementing the MA97 linear solver [http://www.hsl.rl.ac.uk/catalogue]. The optimizations were run on a 16-core (3.20 GHz) Linux workstation with 64 GB of RAM and a University of Bath minicomputer with 64-core processors and 320GB of RAM. Model equations were linearized according to Boole's rule.[30]

*Statistical analysis*

Extreme outliers in the predicted charge data were detected using the ROUT test[79] with the maximum desired false discovery rate, Q set at 0.1%, based on values for the NaT channel. Only 3 outliers were identified out of a total of 138. The corresponding parameters solutions $p^*$ could also be identified by their failure to predict the membrane voltage oscillations over the 2000ms epoch.

Due to the non-gaussian distributions of some of the total predicted charge data, multiple two-tailed Mann-Whitney U[80] rank-sum tests were applied, with multiple comparisons corrected for using the two-stage step-up method of Benjamini, Krieger and Yekutieli, with Q at 1%. Mann-Whitney U values are reported, and multiplicity-corrected significance values (q) are therefore reported for all discoveries. In figures, asterisks are applied based on these q values. GraphPad Prism version 9 was used for all statistical analyses.

**Acknowledgements**

**Funding:** This work was supported by the European Union's Horizon 2020 Future Emerging Technologies Programme under grant 732170.

**Author contributions:** PGM performed all electrophysiological experiments and their statistical analysis; JDT performed all computations estimating model parameters, predicting changes in ionic currents both in rodent and model data; JFRP contributed to the design of electrophysiological experiments and reviewed the manuscript; AN and PGM conceived the work; AN supervised the work and secured funding; AN, PGM and JDT wrote the manuscript.  All authors contributed to the submitted version.

**Competing interest:** The authors declare no competing interests

**Data and materials availability:** Electrophysiological recordings underpinning this study are archived on the open data base https://researchdata.bath.ac.uk

**Ethical statement:** Experiments on rodents were performed under Schedule 1 in accordance with the United Kingdom Scientific Procedures act of 1986.

# Figures and Tables

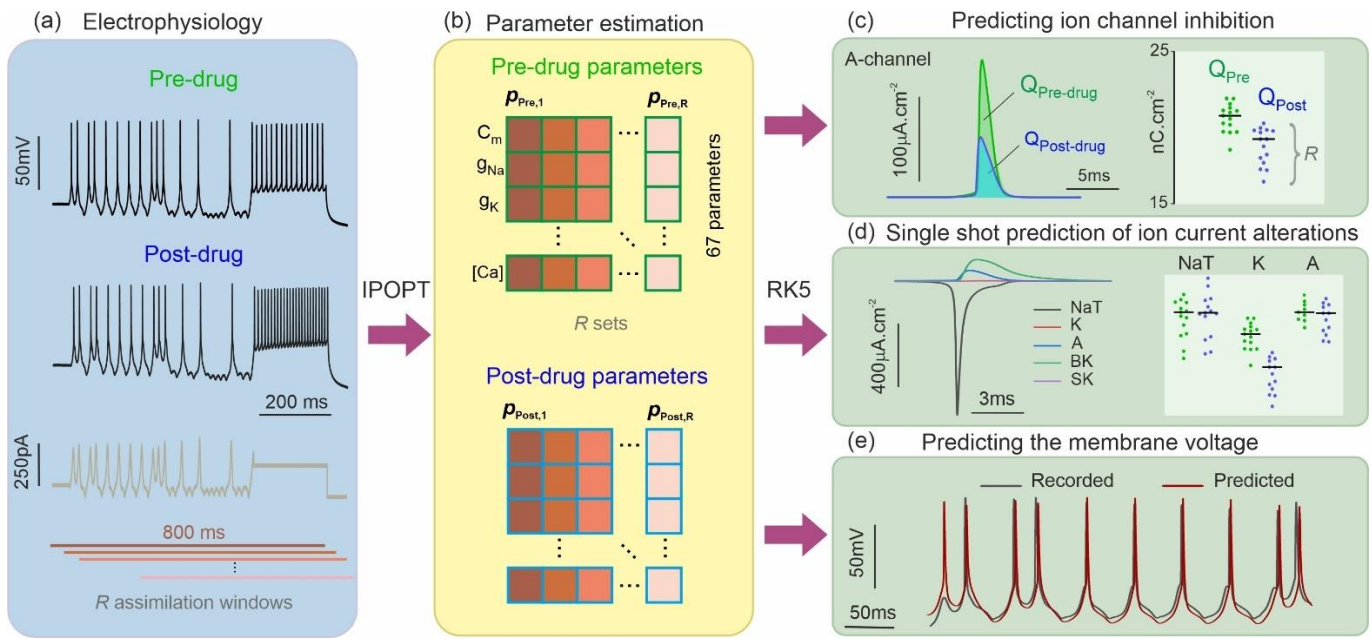

**Figure 1. Estimation of ion current alterations from current clamp recordings**

(a) Membrane voltage of a (hippocampal) neuron recorded before and after a pharmacological inhibitor is applied to partially block a specific ion channel (black traces). The same current protocol (brown trace) is applied to elicit pre-drug and post-drug oscillations. (b) Data assimilation (IPOPT) was used to synchronize a 9-ion channel Hodgkin-Huxley model to the data over a 800ms long time window and obtain one set of pre-drug parameters $\{p^*_{Pre}\}$ and one set of post-drug parameters $\{p^*_{Post}\}$. Each set has $K=67$ parameters. This approach was repeated over $R$ assimilation windows offset by 80ms to generate a statistical sample of parameter sets $\{p^*_{Pre}\}_{1,\ldots,R}$ and $\{p^*_{Post}\}_{1,\ldots,R}$ where $R=15\text{-}19$ depending on the antagonist applied. (c-e) The Hodgkin-Huxley model configured with each set of estimated parameters was used to predict the ionic current waveforms and membrane voltage oscillations through forward integration of the current protocol with an adaptive step-size fifth order Runge-Kutta method (RK5). (c,d) The degree of channel block was predicted by calculating the amount of ionic charge transferred per action potential, $Q_{Pre-drug}$ and $Q_{Post-drug}$, for all 9 ion channels of the model. Predictions were validated by comparing the mean reduction in charge transfer to the known selectivity and potency of the antagonist. (e) Predictions were also validated by comparing the predicted membrane voltage to the measured one.

Page 28 of 34

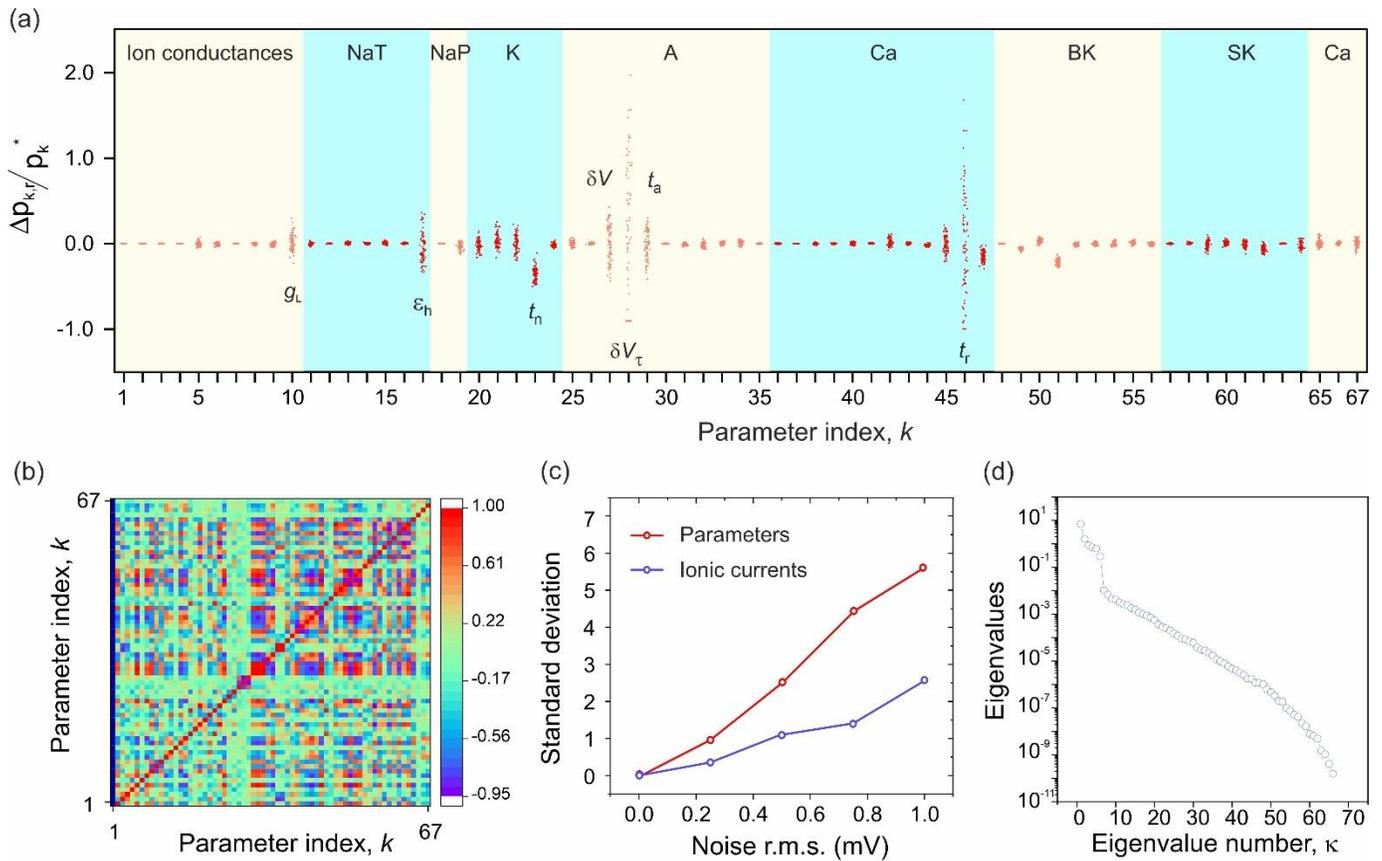

**Figure 2. Comparing the uncertainty on estimated ionic currents and parameters**

(a) Deviations of parameter estimates from their true values ($p_k^*$), $k = 1 \ldots 67$, when Gaussian noise is added to the membrane voltage. Parameter deviations $\Delta p_{k,r} = (p_{k,r} - p_k^*)$ were computed from a statistical sample of $R = 100$ assimilations of the same 800ms window with 100 different realizations of added noise (0.25mV *r.m.s.*) (red dots). The greater the dispersion, the greater the parameter sensitivity to data (and model) error. (b) Covariance matrix of parameter deviations: $\sigma_{kk'} = \frac{1}{R-1}\sum_{r=1}^{R}\left(\frac{\Delta p_{k,r}}{p_k^*}\right)\left(\frac{\Delta p_{k',r}}{p_{k'}^*}\right)$. Correlations occur within blocks of parameters pertaining to the *same ionic current*. In contrast, correlations between the parameters of different ionic currents are weaker. (c) Comparison of the standard deviations of predicted ionic currents and of their underlying parameters. On average over the *r.m.s.* noise amplitudes, 0.25mV, 0.5mV, 075mV, and 1.0mV, the *uncertainty on ionic currents is 3 times smaller than on parameters*. (d) Spectrum of eigenvalues of the covariance matrix. While most parameters are well-constrained the 6 outliers relate to the loosely constrained recovery time constants $t$ and $\epsilon$ in panel *a*.

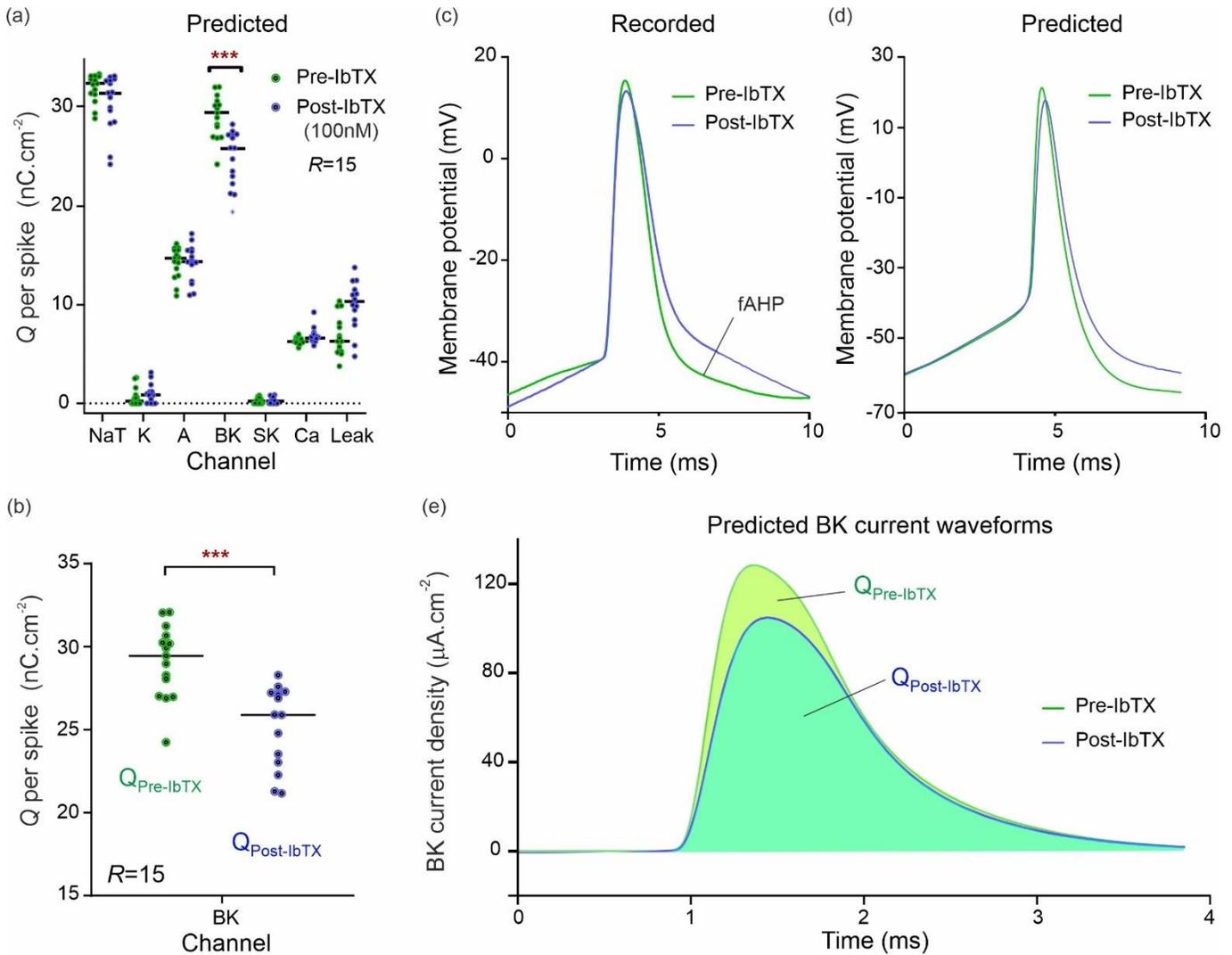

**Figure 3. Single-shot prediction of ionic current block by Iberiotoxin (IbTX)**

(a) Predicted ionic charge transferred per action potential, per ion channel, across the complement of ion channels of a CA1 neuron. The green dots are the charge predictions computed from $R$=15 assimilation windows of pre-drug neuron recordings. The blue dots are the charge predictions computed similarly from the same neuron after 100 nM IbTX was applied. Horizontal bars show median charge values. Asterisks (***) indicate multiplicity adjusted q values from multiple Mann-Whitney U tests using a False Discovery Rate approach of 1%. (b) Predicted change in BK charge transfer showing the effect of IbTX as the nominal BK antagonist. (c) Effect of IbTX measured in one action potential. Inhibition of the BK channels reduces afterhyperpolarization (fAHP). (d) Effect of IbTX predicted for the same action potential. Each voltage trace is the average of 15 waveforms computed from 15 assimilations windows. (e) Predicted BK current waveforms and their alteration by IbTX. Each waveform is the average of 15 BK current waveforms reconstructed from 15 assimilation windows.

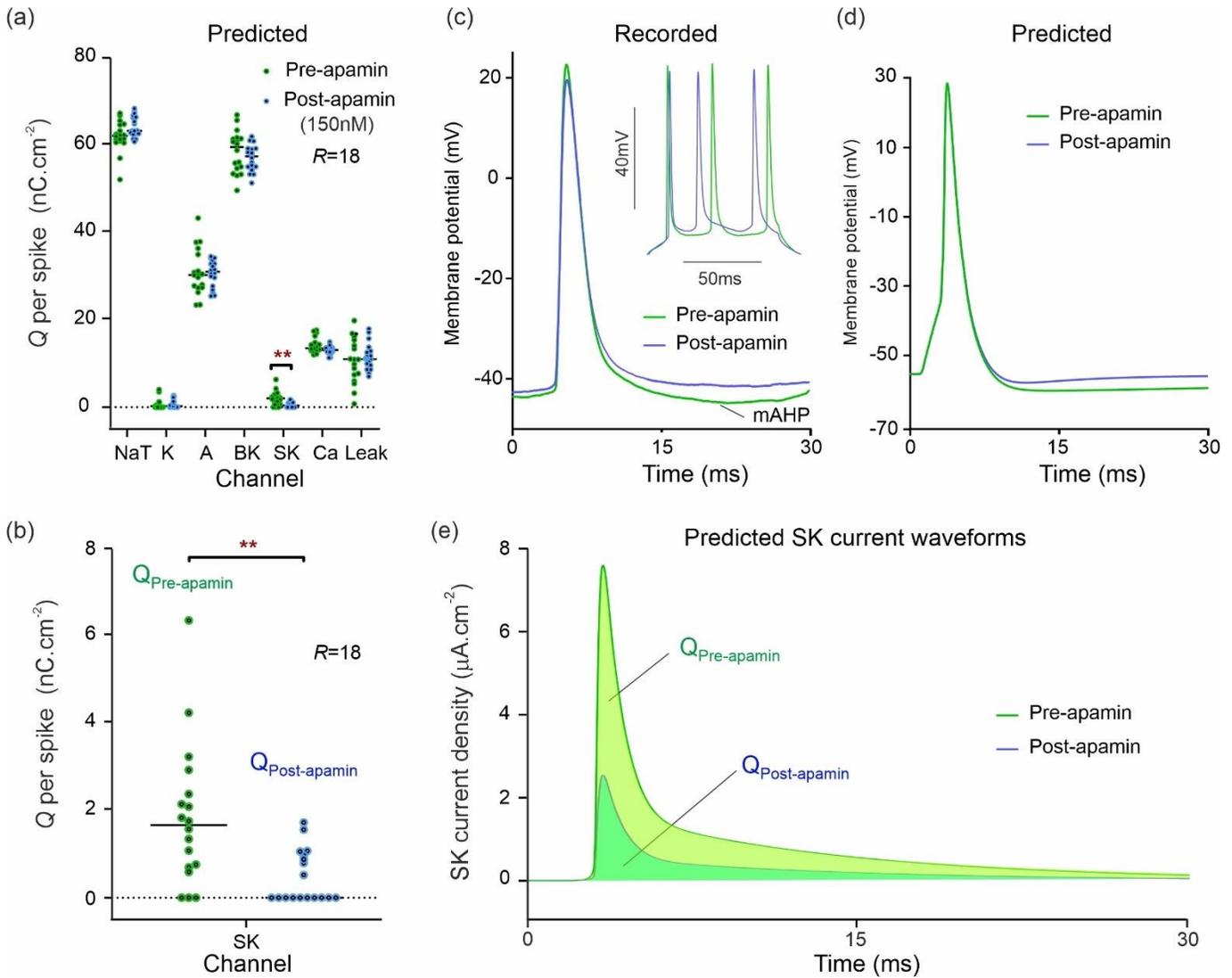

**Figure 4. Single shot prediction of ionic current block by apamin.**

(a) Predicted ionic charge transferred per action potential, per ion channel, of a CA1 neuron. The green dots are the ionic charges predicted from $R$=18 assimilation windows of pre-drug recordings. The blue dots are the predictions computed similarly after 150nM apamin was applied to the neuron. Horizontal bars are the median charge values. Asterisks (**) represent multiplicity adjusted q values from multiple Mann-Whitney U tests using a False Discovery Rate approach of 1%. (b) Predicted change in SK charge transfer showing the effect of apamin as the nominal SK antagonist. (c) Effect of apamin on one action potential and multiple action potentials (*inset*). (d) Effect of apamin predicted for the same action potential. Each voltage trace averages 18 waveforms computed from 18 assimilation windows. (e) Predicted SK current waveforms and their alteration by apamin. Each waveform is the average of 18 current waveforms reconstructed from 18 assimilation windows.

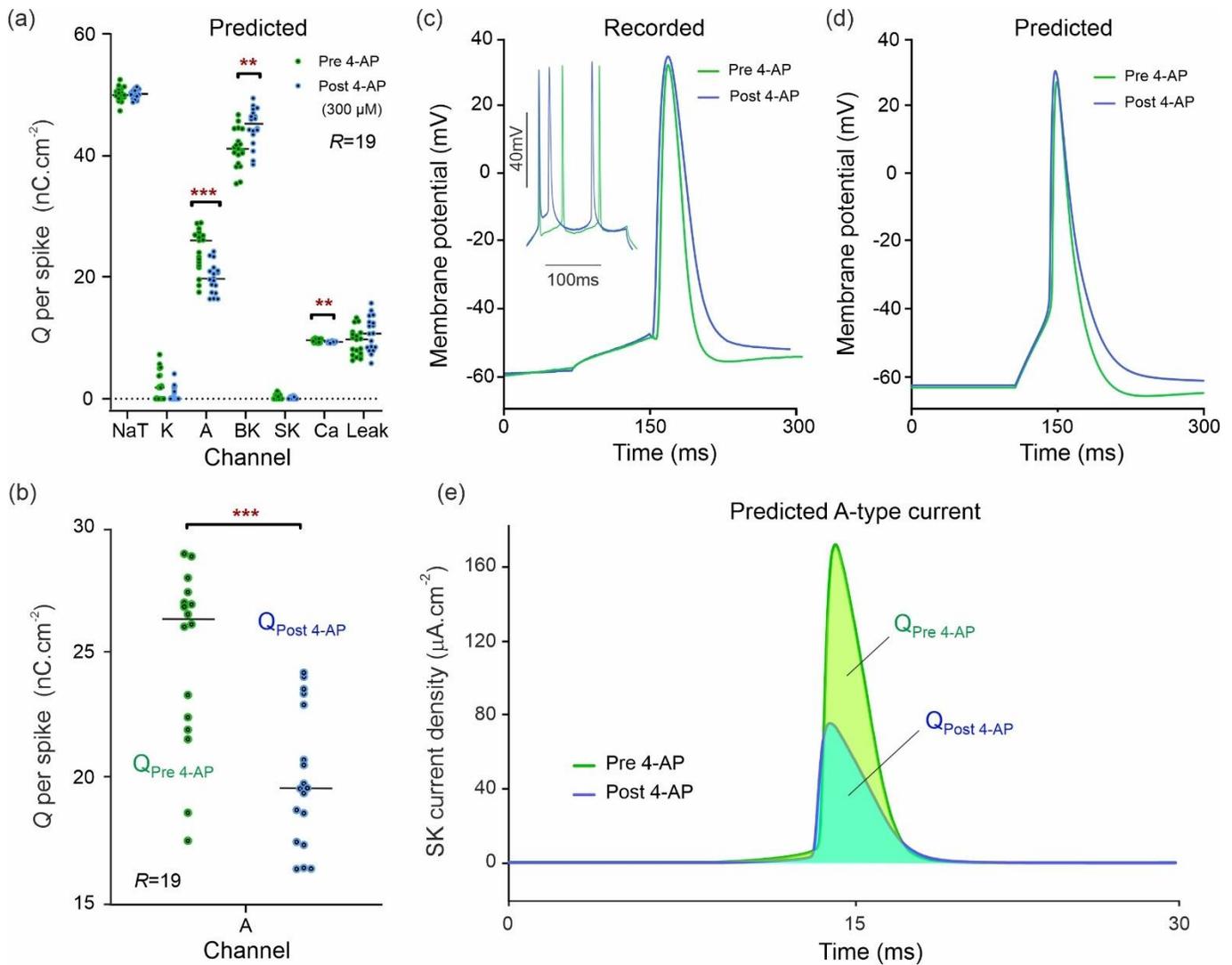

**Figure 5. Single shot prediction of ionic current block by 4-Aminopyridine (4-AP)**

(a) Predicted ionic charge transferred per action potential, per ion channel of a CA1 neuron. The green dots show the charge predicted from $R$=19 assimilation windows of pre-drug recordings. The blue dots show the same after 300 µM 4-AP after was applied to the neuron. (b) Predicted change in A-type charge transfer showing the effect of 4-APP as the nominal A-type antagonist. (c) Effect of 4-AP on one action potential. *Inset*: 4-AP increases the speed of adaptation of the neuron to stimulation following removal of the A-current-mediated delay. (d) Effect of 4-AP predicted for the same action potential. Each trace averages 19 waveforms computed from 19 assimilation windows. (e) Predicted A-type current waveforms and their alteration by 4-AP. Each waveform is the average of 19 waveforms reconstructed from 19 assimilation windows.

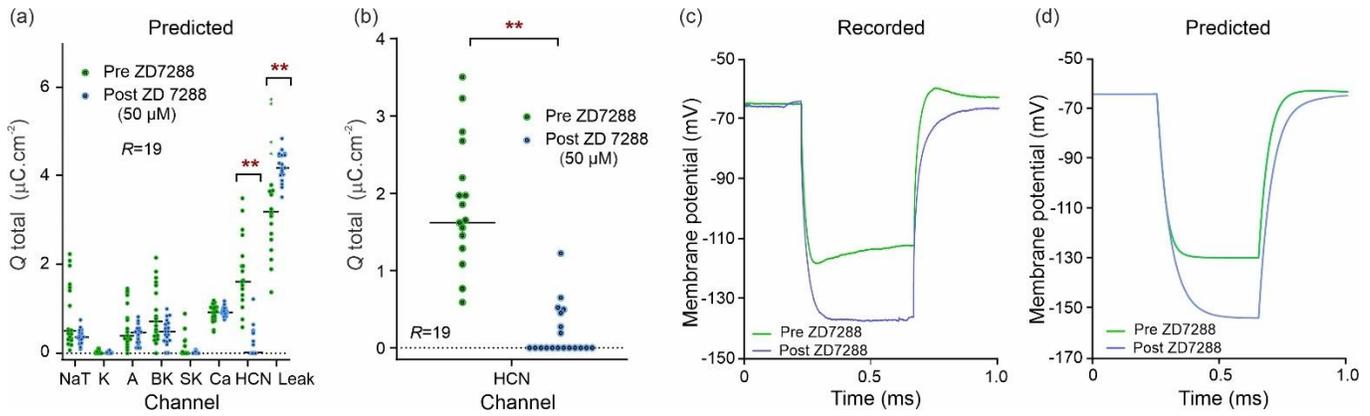

**Figure 6. Single shot prediction of the ionic current block by ZD7288**

(a) Predicted ionic transferred per ion channel over an entire 800ms long assimilation window for a CA1 neuron. The green dots are the charge predicted from $R$=19 assimilations of pre-drug recordings. The blue dots are the charge predictions computed similarly after 50 µM ZD7288 was applied. (b) Predicted change in HCN-type charge transfer showing the effect of ZD7288 as the nominal HCN antagonist. (c) Effect of ZD7288 observed during a hyperpolarizing current step activating the HCN channel. (d) Effect of the ZD7288 antagonist predicted in response to the same hyperpolarizing step. Each waveform is an average of 19 waveforms reconstructed from 19 assimilation windows.

| Drug | Dose [Mol] | Ion channel [sub-unit] | | Expected block [Reference] | | Median & Mean inhibition | CoV |
|---|---|---|---|---|---|---|---|
| IBTX | 0.1 | BK | [α] | 86 % | [43] | 12.1 % & 14.8 % | 21.3 % |
| | | BK | [α+β1] | 12 % | [43] | | |
| | | BK | [α+ β3] | 77 % | [43] | | |
| | | BK | [α+ β4] | Insensitive | [43] | | |
| Apamin | 0.15 | SK | [mAHP] | ~100 % | [49] | 100 % & 74.0 % | 0.1 % |
| | | SK2 | (+++) | ~100 % | [51] | | |
| | | SK3 | (+) | ~80 % | [51] | | |
| | | SK1 | (++) | Insensitive | [50] | | |
| 4-AP | 300 | Kv1.4 | (+) | ~25 % | [56] | 24.3 % & 19.0 % | 21.1 % |
| | | Kv4.2 | (++) | ~13 % | [57] | | |
| ZD7288 | 50 | HCN | | 70 % | [67] | 100 % & 85.5 % | 24.1 % |
| | | | | 85 % | [71] | | |
| | | HCN1 | | ~63 % | [64] | | |
| | | HCN2 | | ~80 % | [66] | | |

**Table 1. Validation of the predicted current alterations against the known potency of antagonists**
The median and mean degrees of inhibition predicted using data assimilation are compared to the known potency of the antagonist at the same concentration as in our experiments. (+)/(++)/(+++) indicates the relative degree of expression (low, medium, high) of individual ion channel sub-types[47–49].

# Supplementary Materials for

## Single shot diagnosis of ion channel dysfunction from assimilation of cell membrane dynamics

Paul G. Morris, Joseph D. Taylor, Julian F.R. Paton, Alain Nogaret*

*Corresponding author. Email: A.R.Nogaret@bath.ac.uk

**This PDF file includes:**

Supplementary Text
Tables S1 to S2
Figures S1 to S3

**Other Supplementary Materials for this manuscript include the following:**

Source data file

**Supplementary text: Computational method and analysis of data**

1. **Completed models successfully predict membrane oscillations over 2s long epochs**

**Figure S1** shows the predictions of the completed model over the full 2000ms long epoch. This epoch covers a wider time interval than the few action potentials of Figures 3-5. The membrane volage predicted by the pre-drug model is compared to the measured membrane voltage time (Panel A) and similarly, the membrane voltage predicted by the post-drug model is compared to the membrane voltage measured after applying apamin (panel B). In each case, the models make excellent predictions of the membrane volage over the 2000ms long epoch. Details are shown in panels C and D.

2. **Calibrating the amplitude of the stimulating current**

In order to determine the range of optimal current stimulation prior to any measurement being taken, we conducted the calibration exercise depicted in **Figure S2a**. This protocol injected 50ms long current steps increasing gradually from 20pA to a maximum of 600pA in 20pA increments. The optimal level of current stimulation maximizes the number of action potentials per epoch while minimizing depolarization block that occurs at higher stimulation. In this way, a maximum amount of information could be transferred from the biological data to the model. Once the optimal level of current stimulation was established, we applied the appropriate scaling factor to each of the 150 protocols generating the assimilation data. **Figure S2b** is an example of a suitably scaled current protocol in contrast to **Figure S2c** which has fewer action potentials when stimulation is too large (depolarization block). Low neuron firing rates occur both at the low end of the stimulation range when the current is near the threshold and at the high end when the neuron undergoes depolarization block.

3. **Ionic charge integrated over the 2s long epoch**

Here we present an alternative method for calculating the charge transfer per ion channel. Instead of calculating the charge transferred per action potential as done in Figures 3-5, we calculate the charge transferred across the entire 2000ms assimilation window through both sub-threshold regions and action potentials (**Figure S3**). Ionic charge transfer during sub-threshold oscillations is expected to be negligible, however this needs to be verified to demonstrate the robustness of the claimed predictions and their independence on the method used. The percentage drop in ionic charge transfer across the 2000ms window is quoted in dark bold **[…%]** against the percentage drop per action potential in green bold font **[…%]**.

*BK channel blockade*

The BK channel blocker, iberiotoxin (IbTX; 100 nM; pre-drug assimilations/predictions $R$=15; post drug $R$=15) reduced the transfer of BK-specific charge by **13.1%** **[12.1%]** (U=28.5; q<0.01) relative to the pre-drug state. The median charge transfer across the 2000ms window was 1.697

µC.cm$^{-2}$ pre-drug, and 1.474 µC/cm$^{-2}$ in IbTX (**Figure S3 A, B**). Data assimilation also predicted an increase in leak current in IbTX (U=41.5; q<0.01; median 0.378 µC.cm$^{-2}$ [pre-drug], 0.599 µC.cm$^{-2}$ [IbTX]). This is a secondary effect caused by the reduction in K$^+$ permeability when inhibiting K$^+$ channels, increasing the driving force of Cl$^-$ into the cell. The mean number of spikes over the 2000ms window was 39.47 ± 0.86 before IbTX, and 38.40 ± 0.855 after IbTX was applied (*R* values as above).

*SK channel blockade*
The SK-specific channel blocker, apamin (150 nM; pre-drug assimilations/predictions *R*=18; post-drug *R*=18), reduced the transfer of SK-specific charge by **100% [100%]** (U=68; q<0.01; median 0.0483 µC.cm$^{-2}$ [pre-drug], 0.000 µC.cm$^{-2}$ [apamin]) relative to the pre-drug state (**Figure S3 C, D**). Data assimilation predicted an additional increase in charge transfer through the voltage-gated Na$^+$ channel (U=56; q<0.01; median -2.017 µC.cm$^{-2}$ [pre-drug], -2.263 µC.cm$^{-2}$ [apamin]) and the A-type K$^+$ channel (U=66; q<0.01; median 0.960 [pre-drug], 1.070 [apamin]).

*Blockade of A-type channels*
The application of 300 µM of the Kv channel blocker 4-AP (pre-drug assimilations/predictions *R*=19; post-drug *R*=18), reduced the transfer of A-type specific charge by **15.5% [19%]** (U=62; q<0.001; medians 0.841 µC.cm$^{-2}$ [pre-drug], 0.710 µC.cm$^{-2}$ [4-AP]) relative to the pre-drug state (**Figure S3 E, F**). Data assimilation predicted an increase in charge transfer through the BK (U=40; q<0.0001; medians 1.394 µC.cm$^{-2}$ [pre-drug], 1.630 µC.cm$^{-2}$ [4-AP]), voltage-gated Na$^+$ (U=59; q<0.001; medians -1.682 µC.cm$^{-2}$ [pre-drug], -1.807 µC.cm$^{-2}$ [4-AP]), and Ca$^{2+}$ (U=83; q=0.01; medians -0.320 µC.cm$^{-2}$ [pre-drug], -0.332 µC.cm$^{-2}$ [4-AP]) channels.

In summary computing the charge transferred per action potential per ion channels give very similar results to computing the charge transferred period channels across the assimilation window for the BK, SK and A-type channels.

| ID | Channel | Current density |
|---|---|---|
| NaT | Transient sodium current | $J_{NaT} = g_{NaT}\, m_\infty^3\, h\, (E_{Na} - V)$ |
| NaP | Persistent sodium current | $J_{NaP} = g_{NaP}\, p_\infty (E_{Na} - V)$ |
| K | Delayed-rectifier potassium current | $J_K = g_K n^4 (E_K - V)$ |
| A | A-type potassium current | $J_A = g_A\, a\, b\, (E_K - V)$ |
| Ca | Calcium current | $J_{Ca} = g_{Ca}\, s^2\, r\, (E_{Ca} - V)$ |
| BK | Large conductance calcium-activated potassium current | $J_{BK} = g_{BK}\, c^2\, d\, (E_K - V)$ |
| SK | Small conductance calcium-activated potassium current | $J_{SK} = g_{SK}\, w\, (E_K - V)$ |
| HCN | Hyperpolarization-activated cation current | $J_{HCN} = g_H\, z\, (E_{HCN} - V)$ |
| Leak | Leakage current | $J_{Leak} = g_L (E_L - V)$ |

**Table S1. The 9 ionic currents of our hippocampal neuron model.**
The voltage dependence of the ionic current densities is shown in the right column. Parameters include ionic conductances $g_{i=\{NaT,NaP,K,A,Ca,BK,SK,HCN,L\}}$, reversal potentials $E_{j=\{Na,K,Ca,HCN,L\}}$ listed in Table S2. The model has 14 gate variables $\{V, m, h, p, n, a, b, s, r, c, d, w, z, [Ca]_{in}\}$ and 67 parameters.

| $i$ | $p_i$ | | Units | LB | UB | $i$ | $p_i$ | | Units | LB | UB |
|---|---|---|---|---|---|---|---|---|---|---|---|
| 1 | Cap. | $C_m$ | µF/cm² | 1 | 1 | 35 | A | $\varepsilon_b$ | ms | 5 | 50 |
| 2 | NaT | $g_{NaT}$ | nS/cm² | 5 | 100 | 36 | Ca | $g_{Ca}$ | nS/cm² | 9 | 12 |
| 3 | NaP | $g_{NaP}$ | nS/cm² | 5 | 100 | 37 | Ca | $E_{Ca}$ | mV | 120 | 120 |
| 4 | Na | $E_{Na}$ | mV | 60 | 70 | 38 | Ca | $V_s$ | mV | -35 | -25 |
| 5 | K | $g_K$ | nS/cm² | 5 | 20 | 39 | Ca | $\delta V_s$ | mV | 10 | 20 |
| 6 | HCN | $g_{HCN}$ | nS/cm² | 0 | 0.3 | 40 | Ca | $\delta V_{\tau s}$ | mV | 30 | 40 |
| 7 | K | $E_K$ | mV | -110 | -90 | 41 | Ca | $t_s$ | ms | 0.01 | 0.1 |
| 8 | Leak | $E_L$ | mV | -75 | -55 | 42 | Ca | $\varepsilon_s$ | ms | 0.1 | 2 |
| 9 | HCN | $E_{HCN}$ | mV | -60 | -40 | 43 | Ca | $V_r$ | mV | -70 | -55 |
| 10 | Leak | $g_L$ | nS/cm² | 0.2 | 1 | 44 | Ca | $\delta V_r$ | mV | -20 | -10 |
| 11 | NaT | $V_m$ | mV | -40 | -25 | 45 | Ca | $\delta V_{\tau r}$ | mV | 20 | 30 |
| 12 | NaT | $\delta V_m$ | mV | 5 | 20 | 46 | Ca | $t_r$ | ms | 0.1 | 1.0 |
| 13 | NaT | $V_h$ | mV | -70 | -50 | 47 | Ca | $\varepsilon_r$ | ms | 1.0 | 10 |
| 14 | NaT | $\delta V_h$ | mV | -30 | -5 | 48 | BK | $g_{BK}$ | nS/cm² | 0 | 100 |
| 15 | NaT | $\delta V_{\tau h}$ | mV | 20 | 40 | 49 | BK | $V_c$ | mV | -20 | -10 |
| 16 | NaT | $t_h$ | ms | 0.1 | 2.0 | 50 | BK | $\delta V_c$ | mV | 5 | 30 |
| 17 | NaT | $\varepsilon_h$ | ms | 5 | 20 | 51 | BK | $\tau_c$ | ms | 1.1 | 1.1 |
| 18 | NaP | $V_p$ | mV | -40 | -25 | 52 | BK | $V_d$ | mV | -60 | -40 |
| 19 | NaP | $\delta V_p$ | mV | 5 | 20 | 53 | BK | $\delta V_d$ | mV | -20 | -5 |
| 20 | K | $V_n$ | mV | -40 | -25 | 54 | BK | $\delta V_{\tau d}$ | mV | 5 | 30 |
| 21 | K | $\delta V_n$ | mV | 5 | 25 | 55 | BK | $t_d$ | ms | 0.1 | 2.0 |
| 22 | K | $\delta V_{\tau n}$ | mV | 5 | 25 | 56 | BK | $\varepsilon_d$ | ms | 1.0 | 20 |
| 23 | K | $t_n$ | ms | 0.1 | 2.0 | 57 | SK | $g_{SK}$ | nS/cm² | 0 | 0.05 |
| 24 | K | $\varepsilon_n$ | Ms | 1 | 10 | 58 | SK | $V_w$ | mV | 0.5 | 1.0 |
| 25 | A | $g_A$ | nS/cm² | 1 | 100 | 59 | SK | $\delta V_w$ | mV | 0.3 | 0.5 |
| 26 | A | $V_a$ | mV | -20 | 5 | 60 | HCN | $V_z$ | mV | -90 | -70 |
| 27 | A | $\delta V_a$ | mV | 5 | 25 | 61 | HCN | $\delta V_z$ | mV | -20 | -1 |
| 28 | A | $\delta V_{\tau a}$ | mV | 5 | 25 | 62 | HCN | $\delta V_{\tau z}$ | mV | 10 | 30 |
| 29 | A | $t_a$ | ms | 0.1 | 2.0 | 63 | HCN | $t_z$ | ms | 1 | 10 |
| 30 | A | $\varepsilon_a$ | ms | 1.0 | 20 | 64 | HCN | $\varepsilon_z$ | ms | 10 | 200 |
| 31 | A | $V_b$ | mV | -90 | -80 | 65 | Area | A | x10⁴ µm² | 1.0 | 5.0 |
| 32 | A | $\delta V_b$ | mV | -20 | -5 | | | | | | |
| 33 | A | $\delta V_{\tau b}$ | mV | 20 | 30 | 66 | Ca | $\tau_{Ca}$ | ms | 1.0 | 2.0 |
| 34 | A | $t_b$ | ms | 5 | 50 | 67 | Ca | $[Ca]_\infty$ | mM | 0.001 | 0.001 |

**Table S2. List of model parameters with their search interval [LB-UB]**
The model parameters and parameter search intervals for the conductance-based model used for assimilating hippocampal neuron data. LB and UB are the lower and upper boundaries of the parameter search interval respectively.

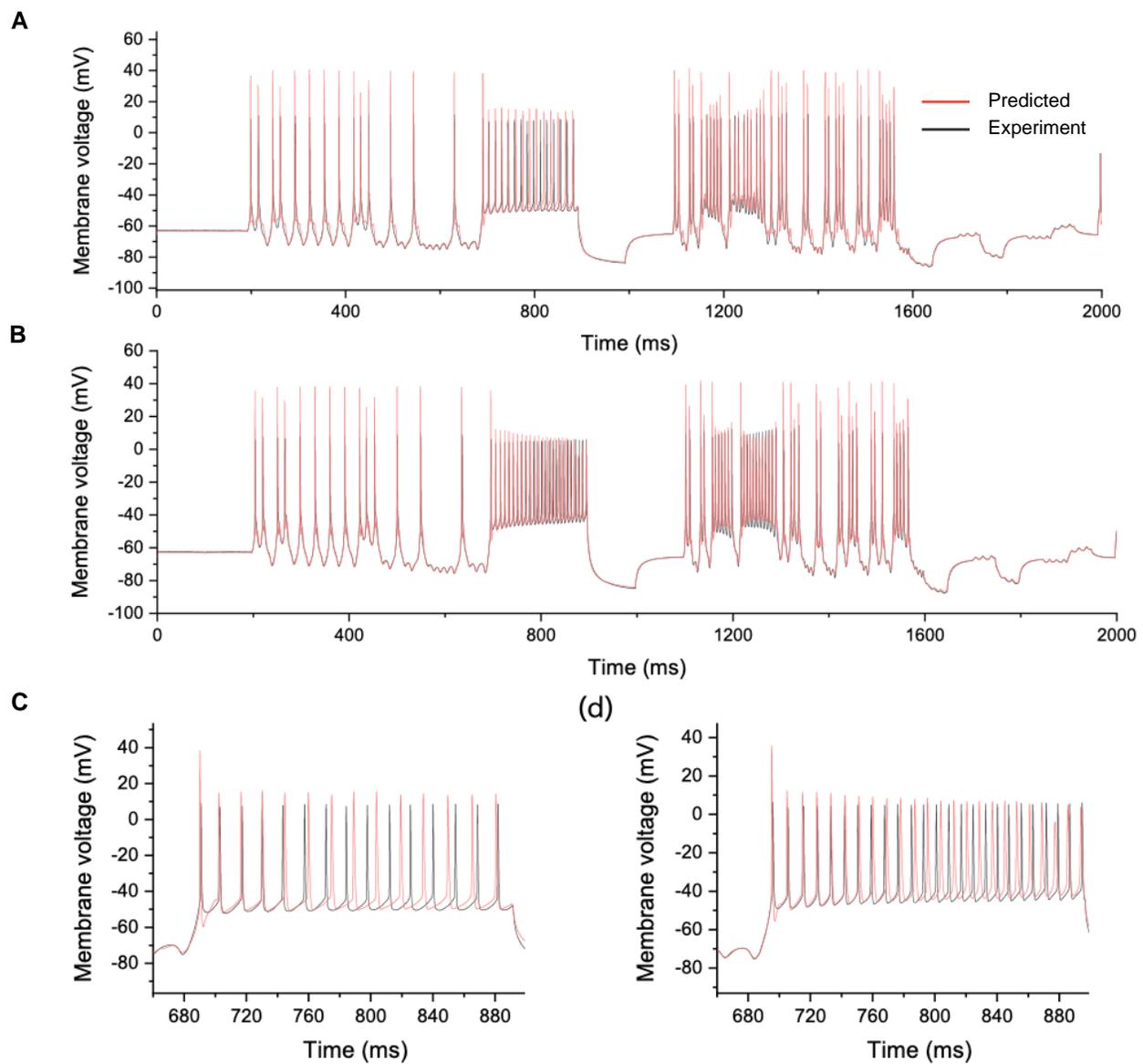

**Figure S1. Long-term prediction of membrane voltage by the completed model before and after the application of apamin (SK blockade)**.
(**A**) 2000-ms-long epoch showing the membrane voltage measured from a CA1 hippocampal neuron (black line) and the predicted membrane voltage (red line). The predicted trace was obtained by forward integrating the completed CA1 neuron model stimulated by the same current protocol as the real neuron. (**B**) 2000-ms-long epoch showing the same neuron after application of SK-channel blocker apamin (150 nM). The predicted membrane voltage was computed from a new completed model. (**C, D**) Detail of the membrane voltage response to the step current at t=680ms pre- and post-channel blockade, respectively. The model successfully replicates the increase in firing frequency after the blocking of the SK potassium channel.

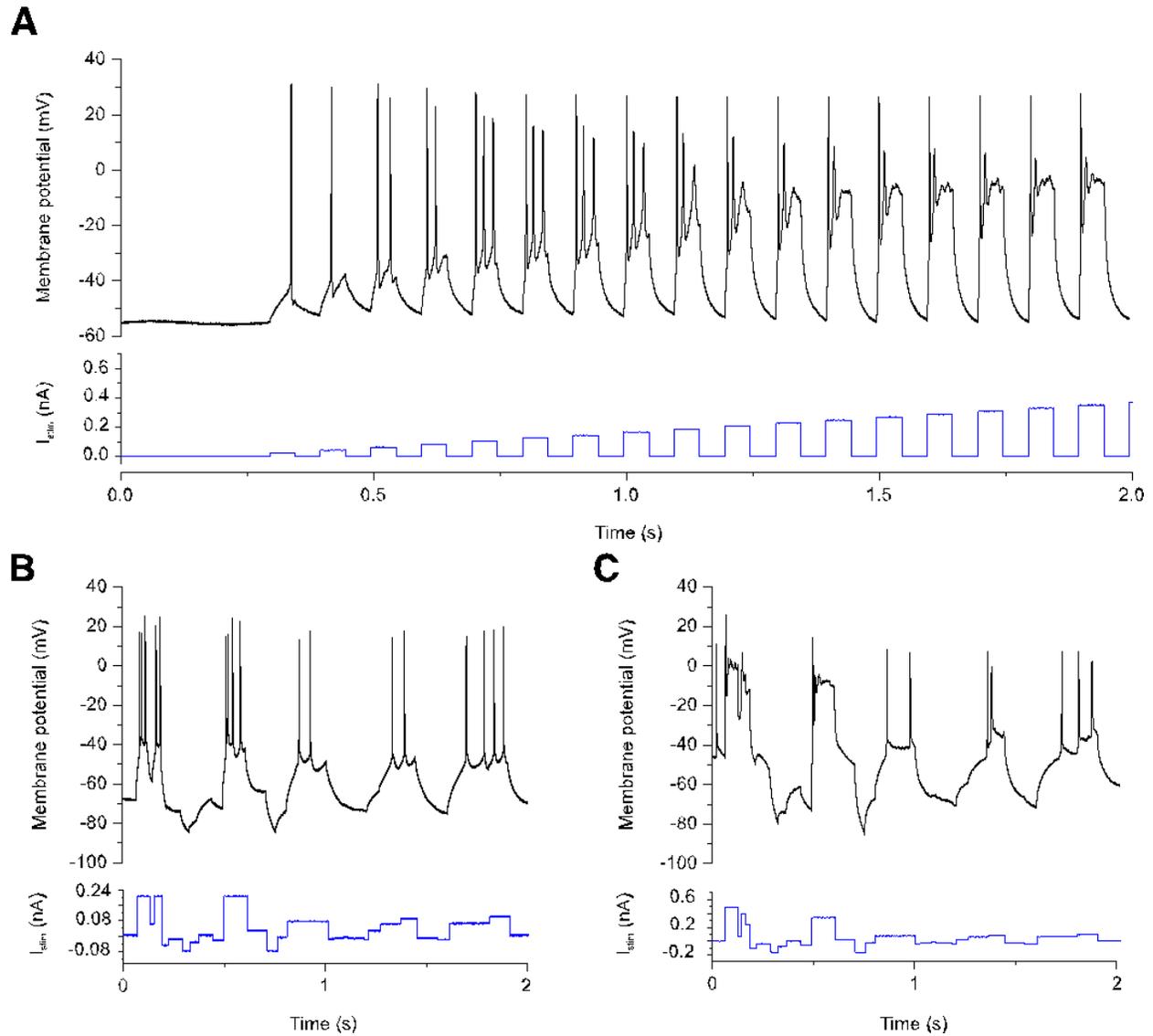

**Figure S2. Action potentials under stimulation by current steps of increasing amplitude**
(**A**) A calibrating current protocol consisting of consecutive step currents of increasing amplitude was used to determine the current range suitable for data assimilation. The amplitudes of current steps in the second half of this particular calibration induce depolarization block, which reduces the amount of information that can be retrieved regarding the underlying ion-channel dynamics. (**B**) A suitably calibrated stimulation protocol using intermediate stimulation currents. (**C**) Exemplar portion of a protocol using large amplitude stimulation currents.

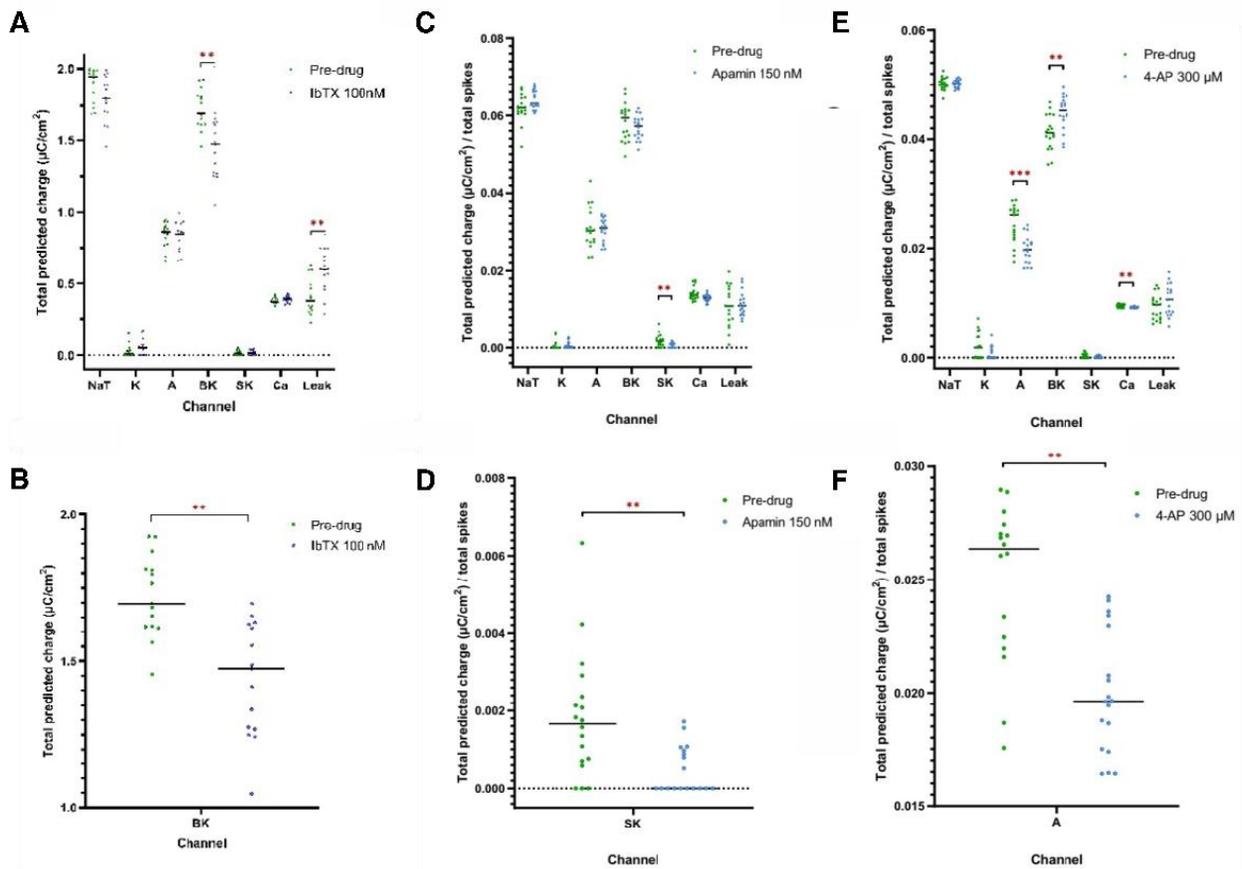

**Figure S3. Total estimated charge transferred per ion channel before and after channel inhibition**. (**A**) Total predicted charge integrated across the 2000ms window of Figs. S1A, S1B for each channel. The green dots ($R$=15) show the charge transferred pre-drug for the NaT, K, SK, BK, A, Ca, Leak channels. The blue dots ($R$=15) show the change in charge transferred after the application of 100nM Iberiotoxin (BK blocker). (**B**) BK detail only. (**C**) Total ion charge transferred across the 2000ms window inferred from pre-drug data (green dots, $R$=18) and after applying 150 nM apamin (SK blocker; blue dots, $R$=18) for all channels. (**D**) SK ion channel detail only. (**E**) Total ion charge transferred across the 2000ms window inferred from pre-drug data ($R$=19), and after 300 µM 4-AP ($R$=18) for all channels. (**F**) A-type channel detail only. Horizontal lines represent median values. Asterisks represent multiplicity adjusted q values from multiple Mann-Whitney U tests using a False Discovery Rate approach (Q) of 1%.